\documentclass[aps,jcp,twocolumn,superscriptaddress,10pt]{revtex4-2}

\usepackage[T1]{fontenc}
\usepackage[utf8]{inputenc}
\usepackage[version=3]{mhchem}
\usepackage{xcolor}
\usepackage{braket}
\usepackage{siunitx}
\usepackage{graphicx}
\usepackage{booktabs}
\usepackage[normalem]{ulem}

\newcommand*{\Scale}[2][4]{\scalebox{#1}{$#2$}}%

\def\al{\ensuremath\alpha}
\def\be{\ensuremath\beta}
\newcommand{\dk}[2]{\delta^{#1}_{#2}}

\begin{document}

\title{First-Order Symmetry-Adapted Perturbation Theory with Double Exchange for Multireference Systems}

\author{Dominik Cieśliński}
\email{d.cieslinski@uw.edu.pl}
\affiliation{University of Warsaw, Faculty of Chemistry, ul.\ L.\ Pasteura 1, 02-093 Warsaw, Poland}

\author{Michał Przybytek}
\affiliation{University of Warsaw, Faculty of Chemistry, ul.\ L.\ Pasteura 1, 02-093 Warsaw, Poland}

\author{Grzegorz Chałasiński}
\affiliation{University of Warsaw, Faculty of Chemistry, ul.\ L.\ Pasteura 1, 02-093 Warsaw, Poland}

\author{Michał Hapka}
\affiliation{University of Warsaw, Faculty of Chemistry, ul.\ L.\ Pasteura 1, 02-093 Warsaw, Poland}

\begin{abstract}
We extend first-order multiconfigurational symmetry-adapted perturbation theory, SAPT(MC), [Hapka M. \textit{JCTC}, 2021, 17], to account for double-exchange effects, where up to two electron pairs are exchanged between interacting monomers. To achieve this, we derive density-matrix-based expressions for the first-order exchange energy to arbitrary orders in the overlap expansion. As a numerical demonstration, we apply the double-exchange approximation to strongly orthogonal geminal wave functions. Additionally, we propose an approximate method for evaluating double-exchange energy with complete active space (CAS) wave functions of the valence type, i.e., with $n$ active electrons distributed over $n$ orbitals. We analyze the performance of these methods on model dimers in both ground and excited states.
\end{abstract}

\keywords{symmetry-adapted perturbation theory, excited states, exchange energy, single-exchange approximation, double-exchange approximation}

\maketitle

\section{Introduction}

The repulsive short-range region of an intermolecular potential is dominated by exchange effects. The exchange phenomenon can be viewed as the tunneling of electrons through the potential barriers between the interacting subsystems. The energy increase when monomer densities overlap, often referred to as Pauli repulsion, arises from the antisymmetry of the electronic wave function of the whole system. Accounting for exchange effects is crucial when modeling nonbonded interactions in force fields \cite{Kuechler:15,VanVleet:16,Rackers:19,Khabibrakhmanov:23}. Exchange also plays an important role at the QM/MM boundary in embedding approaches, where it prevents electron spill-out from the quantum region \cite{Laio:02,Laino:05,Fradelos:11,Reinholdt:17,Giovannini:17,Gokcan:18,Viquez:18,Marefat:20}.

In the perturbation theory, exchange effects appear only after ensuring correct permutational symmetry of the perturbation expansion. Modern symmetry-adapted perturbation theory (SAPT) formulations \cite{Jeziorski:94,Patkowski:20} belong to symmetrized Rayleigh-Schr\"odinger (SRS) \cite{Jeziorski:78} family of methods in which the symmetry forcing operators are applied only in the energy expressions, while preserving the wave function expansion from Rayleigh-Schr{\"o}dinger theory. The resulting exchange energy corrections incorporate an antisymmetrizer, which accounts for all possible exchanges of electrons between interacting monomers. The antisymmetrizer can be expressed as a series of operators that affect $n$ pairs of electrons at a time, enabling a systematic truncation in the energy expressions.

In their seminal work, Jeziorski, Bulski and Piela \cite{Jeziorski:76} demonstrated that the first-order exchange energy can be calculated using the full antisymmetrizer, provided that wave functions of isolated monomers are given as single determinants. More than thirty years later, this approach was generalized for the second order by Sch{\"a}ffer and Jansen \cite{Schaffer:12,Schaffer:13}. In their derivation, the authors used determinants and cofactors of the overlap matrix between orbitals of the monomers. Recently, Waldrop and Patkowski \cite{Waldrop:21} applied this technique to obtain the third-order exchange-induction energy.

Truncation of the antisymmetrizer to the single-exchange operator effectively allows subsystems to interchange only a single pair of electrons. The simplification includes all terms quadratic in elements of the intermolecular overlap matrix $S$, giving rise to the name ``$S^2$ approximation''. Historically, this was the first approach applied to exchange interactions dating back to the second half of the 1960s \cite{Murrell:65,Murrell:67,Salem:65,Musher:66}. Typically, the approximation remains valid at far- and mid-range distances and fails at the repulsive wall where higher-order terms in the overlap expansion can no longer be neglected. The great advantage offered by the single-exchange model is that the first-order exchange energy can be formulated solely in terms of one- and two-electron reduced density matrices (RDMs) of the monomers.~\cite{Moszynski1994a,Korona:08a} For the second-order exchange, additionally one- and two-electron transition reduced density matrices (TRDMs) are required \cite{Korona:08b,Korona:09}. The density-matrix (DM) based SAPT formulation within the $S^2$ approximation was initiated by Moszynski \textit{et al}. \cite{Moszynski1994a} and fully developed by Korona who introduced a coupled-cluster (CC) variant of the method \cite{Korona:08a,Korona:08b,Korona:09,Korona:10} using density matrices derived from the expectation value CC theory (XCC) \cite{Jeziorski:93,Moszynski:05}.

Recently, the DM approach was used to introduce SAPT(MC) \cite{Hapka:19b,Hapka:23}, a multiconfigurational variant of SAPT which extends the applicability of the method to complexes involving strong correlation and excited states. Like its CC-based counterpart, SAPT(MC) incorporates energy corrections through second order in the intermolecular potential and relies on the $S^2$ approximation for exchange. The method is compatible with any multiconfigurational model, provided that one- and two-electron RDMs are available. The necessary TRDMs are obtained from linear response equations in the extended random phase approximation (ERPA) of Pernal and co-workers \cite{Pernal:12,Pernal:14}. To date, SAPT(MC) has proven valuable for analyzing low-lying valence excited states of small organic molecules \cite{Jangrouei:22,Hapka:23}, including a detailed study of paradigm anisole-water and anisole-ammonia complexes \cite{Krzeminska:24}. The method has also been adapted for quantum computing \cite{McClean:16,Loipersberger:23}. In the hybrid approach of Ref.~\cite{Loipersberger:23}, the quantum component of the calculation provides RDMs from active space variational quantum eigensolver wave functions \cite{Peruzzo:14,McClean:16}, while the ERPA response equations are solved classically.

The single-exchange representation of exchange terms is one of the main factors limiting the accuracy of SAPT(MC). The $S^2$ approximation breaks down in systems with diffuse electron densities,
where higher-order overlap contributions become significant even at intermediate intermonomer separations. For example, in anion-neutral and anion-cation complexes studied by Lao et al. \cite{Lao:15}, first-order exchange and second-order exchange-induction energies at the $S^2$ level each exhibit mean absolute errors of the order of 1~kcal/mol. A similar challenge arises in systems with lone electron pairs, such as ammonia interacting with alkali-metal and alkaline-earth-metal atoms \cite{Zuchowski:08}. Slow convergence of the overlap expansion is also expected in potential applications of SAPT(MC), including interactions between molecules in excited states and studies on complexes involving metal-containing compounds.

The first step beyond the $S^2$ model is to account for double exchange effects, i.e., allowing monomers to exchange up to two pairs of electrons, which yields an exchange energy that is exact through the fourth power of the overlap integrals. The pioneering application of the $S^4$ approximation in the first-order perturbation theory was a helium dimer study by Van Duijneveldt-van de Rijdt and Van Duijneveldt in 1972 \cite{Duijneveldt:72}. However, this approach was practically abandoned following the development of the infinite-order expansion by Jeziorski et al.~\cite{Jeziorski:76} just four years later. Only recently have Tyrcha et al.~\cite{Tyrcha:24} revisited the accuracy of the double exchange model for many electron systems through second order in the SAPT expansion. Their results demonstrate that incorporating double exchange leads to a remarkable improvement over the $S^2$ approximation. In particular, in the first order, the $S^4$ approximation reduces the relative percent errors by up to four orders of magnitude.

In this work, we present a general DM formulation of the first-order exchange energy in the $S^4$ approximation. The resulting DM equations can be evaluated with any method that provides one-, two- and three-electron RDMs, and are straightforward to incorporate into the SAPT(MC) framework. Furthermore, inspired by Ref.~\citenum{Tyrcha:24}, we derive
the explicit (without the use of RDMs) first-order exchange energy expression at the $S^4$ level formalism for strongly orthogonal geminal wave functions: antisymmetrized product of strongly orthogonal geminals (APSG) \cite{Kutzelnigg:64} and generalized valence bond perfect-pairing (GVB-PP) \cite{Hurley:53}. This approach, also referred to as the second-quantized (SQ) formulation \cite{Moszynski1994b}, can also be applied to complete active space (CAS) wave functions of the CAS($n,n$) type ($n$ active electrons on $n$ active orbitals), leveraging the approximate mapping between GVB-PP and CAS($n,n$) wave functions \cite{vanMeer:18,Hapka:20}.

We assess the accuracy of the double exchange approximation to the first-order exchange energy computed with APSG and GVB-PP monomer wave functions against SAPT(FCI) benchmarks for model few-electron systems: the \ce{H2}$\cdots$\ce{H2} dimer in out-of-equilibrium geometries, the He$\cdots$\ce{H2} dimer with the hydrogen molecule in either the ground ${}^1\Sigma^+_g$ state or the excited ${}^1\Sigma^+_u$ state, and the beryllium dimer. The benzene-methanethial interaction highlights the importance of effects beyond second order in the overlap for low-lying excited states of both valence and Rydberg character. Using the Be$\cdots$\ce{NH3} dimer and the water dimer as examples, we illustrate the differences between single- and multireference descriptions of first-order exchange effects within the $S^4$ approximation.

\section{Theory}
Consider a weakly interacting dimer $A \cdots B$, in which monomers $A$ and $B$ have $N_A$ and $N_B$ electrons, respectively. The Hamiltonian of the whole system $\hat{H}$ can be split into the sum of Hamiltonians of the individual monomers $\hat{H}_0=\hat{H}_A+\hat{H}_B$, and the interaction operator $\hat{V}=\hat{H}-\hat{H}_0$. In SAPT, we treat the interaction between molecules as a perturbation. The zeroth-order wave function is a product of the wave functions of the individual monomers, $\Psi_A\Psi_B$. In the SRS \cite{Jeziorski:78} formulation of SAPT, one accounts for the proper permutational symmetry by incorporating the antisymmetrization operator $\hat{\mathcal{A}}$ into energy expressions. Since the product wave function $\Psi_A\Psi_B$ already has the correct symmetry with respect to permuations of electrons within each of the monomers, we can write the antisymmetrizer as
\begin{equation}\label{Antisymmetrizer}
 \hat{\mathcal{A}}=\frac{N_{A}!N_{B}!}{\left(N_A+N_B\right)!} \hat{\mathcal{A}}_A\hat{ \mathcal{A}}_B(1+\hat{\mathcal{P}}),
\end{equation}
where $\hat{\mathcal{P}}$ collects all possible exchanges of electron between $A$ and $B$
\begin{equation}
\hat{\mathcal{P}}=\sum_{k=1}^{\Scale[0.6]{{\min(N_A,N_B)}}} \hat{\mathcal{P}}_{2 k}. 
\end{equation}
Operator $\hat{\mathcal{P}}_{2k}$ is a sum of all unique ${N_A\choose k}{N_B\choose k}$ permutation operators that exchange $k$ electron indices in $\Psi_A$ with $k$ electron indices in $\Psi_B$, multiplied by a proper sign factor $(-1)^k$. The first-order interaction energy in SRS is calculated as
\begin{equation} \label{e1full}
    E^{(1)}_{\text{int}}=\frac{\langle\hat{V}\hat{\mathcal{A}} \rangle }{\langle\hat{\mathcal{A}} \rangle},
\end{equation}
where we use the notation $\braket{\hat{X}} = \braket{\Psi_A\Psi_B|\hat{X} |\Psi_A\Psi_B}$. Using the definition of $\hat{\mathcal{A}}$ from Eq.~\eqref{Antisymmetrizer}, we can rewrite Eq.~\eqref{e1full} as
\begin{equation}\label{int}
        E^{(1)}_{\text{int}}=\frac{\langle\hat{V} \rangle+\langle\hat{V}\hat{\mathcal{P}} \rangle }{1+\langle\hat{\mathcal{P}} \rangle} .
\end{equation}
The first-order interaction energy can be expressed as a sum of terms proportional to even powers of the intermolecular overlap integrals. The term proportional to $S^0$ is the familiar electrostatic energy,  $E^{(1)}_{\text{elst}}$, while the remaining terms sum up to the exchange energy, $E^{(1)}_{\text{exch}}$. The individual contribution to the exchange energy proportional to $S^{2n}$ will be denoted as $E^{(1)}_{\text{exch}}(\propto S^{2n})$. Using the Taylor expansion of $(1+x)^{-1}$ and the fact that the expectation value containing $\hat{\mathcal{P}}_{2k}$ is proportional to $ S^{2k}$, we can write the formulas for $E^{(1)}_{\text{elst}}$ and $E^{(1)}_{\text{exch}}(\propto S^{2n})$ for any $n$. Then, the electrostatic energy is given simply as the expectation value of the perturbation operator, $E^{(1)}_{\text{elst}}=\langle \hat{V} \rangle$. The general expression for $E^{(1)}_{\text{exch}}(\propto S^{2n})$ can be written as
\onecolumngrid
\begin{align}\label{E1exch2k}
  E^{(1)}_{\text{exch}}(\propto S^{2n}) &= \langle \hat{V}\hat{\mathcal{P}}_{2n} \rangle
  + \sum_{s=1}^{n} (-1)^{s} \sum_{\substack{l_1,\ldots,l_s\geq 1\\\sum_{i=1}^sl_i \leq n}} \left[ \langle \hat{V} \hat{\mathcal{P}}_{2(n-\sum_{i=1}^sl_i)}\rangle \prod_{i=1}^{s}\langle\hat{\mathcal{P}}_{2l_i} \rangle \right],
\end{align}
\twocolumngrid
where the $\langle \hat{V}\hat{\mathcal{P}}_0\rangle$ term, when it appears on the right hand side of this equation, should be replaced by $\langle \hat{V}\rangle$. In particular, the exchange energy term proportional to $S^2$ takes the form
\begin{equation}\label{Es2}
    E^{(1)}_{\text{exch}}(\propto S^{2})=\langle \hat{V}\hat{\mathcal{P}}_2 \rangle - \langle \hat{V}\rangle\langle\hat{\mathcal{P}}_2 \rangle,
\end{equation}
and the formula for the exchange energy part proportional to $S^4$ reads
\begin{equation}\label{Es4}
    E^{(1)}_{\text{exch}}(\propto S^{4})=\langle \hat{V} \hat{\mathcal{P}}_4\rangle -\langle \hat{V} \hat{\mathcal{P}}_2\rangle\langle \hat{\mathcal{P}}_2 \rangle- \langle \hat{V} \rangle \langle \hat{\mathcal{P}}_4\rangle+\langle \hat{V} \rangle\langle \hat{\mathcal{P}}_2 \rangle^2.
\end{equation}
Eq.~\eqref{Es2} is known as the \emph{single-exchange} or $S^2$ approximation to the exchange energy, whereas the sum
$E^{(1)}_{\text{exch}}(\propto S^{2}) +  E^{(1)}_{\text{exch}}(\propto S^{4})$ is called \emph{double-exchange} or $S^4$ approximation, which we denote as $E^{(1)}_{\rm exch}(S^4)$. When monomers are described by single Slater determinants, Eq.~\eqref{E1exch2k} reduces to:
\begin{equation}
E^{(1)}_{\text{exch}}(\propto S^{2n}) = \langle \hat{V} \hat{\mathcal{P}}_{2n} \rangle_L,
\end{equation}
where the $L$ subscript denotes the linked-diagrams contribution to a given term. This property of the exchange energy has only recently been formulated as a conjecture by Tyrcha et al.~\cite{Tyrcha:24}

Our aim is to arrive at the first-order exchange energy expression in the double exchange approximation which can be evaluated from density matrices of the monomers, and thus can be integrated in the multiconfigurational SAPT approach in a straighforward manner. To achieve this, we start by deriving the DM formulation of the first-order exchange energy to arbitrary order in the overlap expansion. In other words, we extend the DM approach of Moszynski et al.~\cite{Moszynski1994a} who considered only the $S^2$ truncation of the overlap expansion. The alternative to the DM exchange energy formulation is the SQ approach, also proposed by Moszynski and et al.~\cite{Moszynski1994b} The SQ formalism has recently been generalized to infinite order in the overlap expansion by Tyrcha et al.\cite{Tyrcha:24}, yet their approach is limited to single-determinantal wave functions.

Consider two sets of creation/anihilation operators $a^\dagger_p$, $a_q$ and $b_r^\dagger$, $b_s$ acting on Fock spaces $\mathcal{F}_A$ and $\mathcal{F}_B$ of monomers $A$ and $B$, respectively. The operators obey the standard anticommutation relations in spaces $\mathcal{F}_A$ and $\mathcal{F}_B$; moreover, operators $a$ and $b$ commute:
\begin{equation}
\begin{split}
&\left\{a_p^{\dagger}, a_q^{\dagger}\right\}=  \left\{b_r^{\dagger}, b_s^{\dagger}\right\} = \left\{a_p, a_q\right\}=  \left\{b_r, b_s\right\}=0, \\ & \left\{a_p^{\dagger}, a_q\right\}=\delta_{p q}, \qquad  \left\{b_r^{\dagger}, b_s\right\}=\delta_{r s}, \\ &
\left[a_p^{\dagger}, b_r^{\dagger}\right]=\left[a_p^{\dagger}, b_s\right]=\left[a_q, b_r^{\dagger}\right]=\left[a_q, b_s\right]=0
.
\end{split}
\end{equation}
We introduce a simplified notation for a product of creation and annihilation operators:
\begin{equation}
\begin{split}
 & a^{p_1p_2\ldots p_k}_{q_1q_2\ldots q_k}=a^\dagger_{p_1}\ldots a^\dagger_{p_k}a_{q_k}\ldots a_{q_1}, \\
 & b^{r_1r_2\ldots r_k}_{s_1s_2 \ldots s_k}=b_{r_1}^\dagger\ldots b_{r_k}^\dagger b_{s_k}\ldots b_{s_1} .
\end{split}
\end{equation}

All operators which appear in the first-order SAPT energy expressions can be written in the second-quantized form. The interaction operator $\hat{V}$ is given as 
\begin{equation}
      \hat{V} = V_{AB} + \sum_{rs\in B} (v^A)^s_rb^r_{s}+\sum_{pq\in A} (v^B)^q_pa^p_q+\sum_{\substack{pq\in A\\ rs \in B}}v^{qs}_{pr}a^{p}_q b^{r}_s,
\end{equation}
where $V_{AB}$ is the nuclear repulsion term, $v^A$ and $v^B$ denote interaction of electrons with nuclei of monomer $A$ and $B$, respectively, and $v$ represents the electron-electron interaction. 
It is convenient to introduce the modified interaction potential matrix
\begin{equation}
\tilde{v}^{qs}_{pr}=v^{qs}_{pr}+\delta^q_{p}\frac{(v^A)^{s}_{r}}{N_A}+\frac{(v^B)^{q}_{p}}{N_B}\delta^{s}_{r}+\frac{V_{AB}}{N_AN_B}\delta^q_p\delta^s_r,
\end{equation}
so that the interaction operator becomes
\begin{equation}\label{eq:vgen}
    \hat{V} = \sum_{\substack{pq\in A\\ rs \in B}} \tilde{v}^{qs}_{pr} a^p_q b^r_s.
\end{equation}
The generalized interaction potential matrix can be further extended to arbitrary indices $\kappa,\lambda,\mu,\nu$:
\begin{equation}
\tilde{v}^{\lambda \nu}_{\kappa \mu} = v^{\lambda \nu}_{\kappa \mu} + S^\lambda_{\kappa} \frac{(v^A)^{\nu}_{\mu}}{N_A} + \frac{(v^B)^{\lambda}_{\kappa}}{N_B} S^{\nu}_{\mu} + \frac{V_{AB}}{N_A N_B} S^\lambda_\kappa S^\nu_\mu .
\end{equation}

A general expression for the exchange operators $\hat{\mathcal{P}}_{2k}$ for $k=1,2,\ldots$ was first presented by Tyrcha et al.~\cite{Tyrcha:24}
\onecolumngrid
\begin{equation}\label{eq:p2k}
    \hat{\mathcal{P}}_{2k}=(-1)^{k}\left(\frac{1}{k!}\right)^2\sum _{\substack{p_1,p_2,\ldots,p_k\\q_1,q_2,\ldots,q_k}}\sum _{\substack{r_1,r_2,\ldots,r_k\\s_1,s_2,\ldots,s_k}}\left[\prod_{i=1}^k S^{s_i}_{p_i}S^{q_i}_{r_i}\right]a^{p_1p_2\ldots p_k}_{q_1q_2\ldots q_k}b^{r_1r_2\ldots r_k}_{s_1s_2 \ldots s_k}.
\end{equation}
From the definitions in Eq.~\eqref{eq:vgen} and Eq.~\eqref{eq:p2k} we can rewrite $\hat{V}\hat{\mathcal{P}}_{2k}$ as a sum of operators in the normal form
\begin{align}\label{normalVP}
    \hat{V}\hat{\mathcal{P}}_{2k} &=(-1)^{k} \left(\frac{1}{k!}\right)^2\sum _{\substack{p,p_1,p_2,\ldots,p_k\\q,q_1,q_2,\ldots,q_k}}\sum _{\substack{r,r_1,r_2,\ldots,r_k\\s,s_1,s_2,\ldots,s_k}}\left[ \tilde{v}^{qs}_{pr}\prod_{i=1}^kS^{s_i}_{p_i}S^{q_i}_{r_i}\right]a^{pp_1p_2\ldots p_k}_{qq_1q_2\ldots q_k}b^{rr_1r_2\ldots r_k}_{ss_1s_2 \ldots s_k} \\ \nonumber
    &+ (-1)^{k} \frac{1}{k!(k-1)!}\sum _{\substack{p,p_2,\ldots,p_k\\q_1,q_2,\ldots,q_k}}\sum _{\substack{r,r_1,r_2,\ldots,r_k\\s,s_1,s_2,\ldots,s_k}}\left[\tilde{v}^{s_1s}_{pr}S^{q_1}_{r_1}\prod_{i=2}^kS^{s_i}_{p_i}S^{q_i}_{r_i}\right]a^{pp_2\ldots p_k}_{q_1q_2\ldots q_k}b^{rr_1r_2\ldots r_k}_{ss_1s_2 \ldots s_k}\\ \nonumber
    & +(-1)^{k} \frac{1}{(k-1)!k!} \sum_{\substack{p,p_1,p_2,\ldots,p_k\\q,q_1,q_2,\ldots,q_k}} \sum_{\substack{r,r_2,\ldots,r_k\\s_1,s_2,\ldots,s_k}}\left[\tilde{v}^{qq_1}_{pr}S^{s_1}_{p_1}\prod_{i=2}^kS^{s_i}_{p_i}S^{q_i}_{r_i}\right]a^{pp_1p_2\ldots p_k}_{qq_1q_2\ldots q_k}b^{rr_2\ldots r_k}_{s_1s_2 \ldots s_k}\\ \nonumber
    &+(-1)^{k} \left(\frac{1}{(k-1)!}\right)^2\sum _{\substack{p,p_2,\ldots,p_k\\q_1,q_2,\ldots,q_k}}\sum _{\substack{r,r_2,\ldots,r_k\\s_1,s_2,\ldots,s_k}}\left[\tilde{v}^{s_1q_1}_{pr}\prod_{i=2}^kS^{s_i}_{p_i}S^{q_i}_{r_i}\right]a^{pp_2\ldots p_k}_{q_1q_2\ldots q_k}b^{rr_2\ldots r_k}_{s_1s_2 \ldots s_k} .
\end{align}
In the derivation we used the fact that for $x=a,b$:
\begin{equation}\label{commut}
    x^{\kappa}_\lambda x^{\kappa_1\kappa_2\ldots \kappa_k}_{\lambda_1 \lambda_2 \ldots \lambda_k}=x^{\kappa \kappa_1\kappa_2\ldots \kappa_k}_{\lambda \lambda_1 \lambda_2 \ldots \lambda_k}+\sum_{i=1}^{k}\delta_{\lambda}^{\kappa_i}x^{\kappa_1 \kappa_2\ldots \kappa_{i-1} \kappa \kappa_{i+1}\ldots \kappa_k}_{\lambda_1 \lambda_2\ldots \lambda_{i-1}\lambda_i \lambda_{i+1}\ldots \lambda_k} ,
\end{equation}
and
\begin{equation}\label{RI}
\sum_{q \in A}\tilde{v}^{qs}_{pr}S_q^{s_i}=\tilde{v}^{s_is}_{pr}, \quad \sum_{s \in B}\tilde{v}^{qs}_{pr}S_s^{q_j}=\tilde{v}^{qq_j}_{pr}, \quad \sum_{\substack{q \in A\\s \in B}}\tilde{v}^{qs}_{pr} S_{q}^{s_i} S_s^{q_j} = \tilde{v}^{s_iq_j}_{pr},
\end{equation}
resulting from the resolution of identity.

To calculate the expectation value $\braket{\hat{V}\hat{\mathcal{P}}_{2k}}$, recall the definition of a $k$-body reduced density matrix ($k$-RDM) \cite{McWeeny:59} for monomer $X=A,B$
\begin{equation}
\left(\Gamma_{X}^{(k)}\right)^{\kappa_1,\kappa_2,\ldots,\kappa_k}_{\lambda_1,\lambda_2,\ldots,\lambda_k}=\langle \Psi_X \left| x^{\kappa_1,\kappa_2,\ldots,\kappa_k}_{\lambda_1,\lambda_2,\ldots,\lambda_k} \right| \Psi_X \rangle.
\end{equation}
By introducing intermediates similar to the ones proposed by Moszynski et al.~\cite{Moszynski1994a}
\begin{align} \label{intermed}
   (G_k)^{pr}_{qs}&= \left(\frac{1}{k!}\right)^2\sum _{\substack{p_1,p_2,\ldots,p_k\\q_1,q_2,\ldots,q_k}}\sum _{\substack{r_1,r_2,\ldots,r_k\\s_1,s_2,\ldots,s_k}}\left[\prod_{i=1}^kS^{s_i}_{p_i}S^{q_i}_{r_i}\right]\left(\Gamma_{A}^{(k+1)}\right)^{pp_1p_2\ldots p_k}_{qq_1q_2\ldots q_k}\left(\Gamma_{B}^{(k+1)}\right)^{rr_1r_2\ldots r_k}_{ss_1s_2 \ldots s_k}, \\  (F_k)^{pr}_{s_1s}&=\frac{1}{k!(k-1)!}\sum _{\substack{p_2,\ldots,p_k\\q_1,q_2,\ldots,q_k}}\sum _{\substack{r_1,r_2,\ldots,r_k\\s_2,\ldots,s_k}}\left[S^{q_1}_{r_1}\prod_{i=2}^kS^{s_i}_{p_i}S^{q_i}_{r_i}\right]\left(\Gamma_{A}^{(k)}\right)^{pp_2\ldots p_k}_{q_1q_2\ldots q_k}\left(\Gamma_{B}^{(k+1)}\right)^{rr_1r_2\ldots r_k}_{ss_1s_2 \ldots s_k},
 \\  (D_k)^{pr}_{qq_1}&= \frac{1}{(k-1)!k!}\sum _{\substack{p_1,p_2,\ldots,p_k\\q_2,\ldots,q_k}}\sum _{\substack{r_2,\ldots,r_k\\s_1,s_2,\ldots,s_k}}\left[S^{s_1}_{p_1}\prod_{i=2}^kS^{s_i}_{p_i}S^{q_i}_{r_i}\right]\left(\Gamma_{A}^{(k+1)}\right)^{pp_1p_2\ldots p_k}_{qq_1q_2\ldots q_k}\left(\Gamma_{B}^{(k)}\right)^{rr_2\ldots r_k}_{s_1s_2 \ldots s_k}, \\ \label{cinter}
 (C_k)^{pr}_{s_1q_1}&= \left(\frac{1}{(k-1)!}\right)^2\sum _{\substack{p_2,\ldots,p_k\\q_2,\ldots,q_k}}\sum _{\substack{r_2,\ldots,r_k\\s_2,\ldots,s_k}}\left[\prod_{i=2}^k S^{s_i}_{p_i}S^{q_i}_{r_i}\right]\left(\Gamma_{A}^{(k)}\right)^{pp_2\ldots p_k}_{q_1q_2\ldots q_k}\left(\Gamma_{B}^{(k)}\right)^{rr_2\ldots r_k}_{s_1s_2 \ldots s_k},
\end{align}
we arrive at the DM expression for $\langle \hat{V}\hat{\mathcal{P}}_{2k} \rangle$
\begin{equation}\label{dmvp2k}
 \langle \hat{V}\hat{\mathcal{P}}_{2k} \rangle =(-1)^{k}\left(\sum_{\substack{pq\in A\\rs \in B}} (G_k)^{pr}_{qs}\tilde{v}^{qs}_{pr} + \sum_{\substack{p\in A \\ rss_1 \in B}}(F_k)^{pr}_{s_1s}\tilde{v}^{s_1s}_{pr}+ \sum_{\substack{pqq_1\in A \\ r \in B}}(D_k)^{pr}_{qq_1}\tilde{v}^{qq_1}_{pr} +\sum_{\substack{pq_1\in A \\ rs_1 \in B}}(C_k)^{pr}_{s_1q_1}\tilde{v}^{s_1q_1}_{pr} \right).
\end{equation}
For completeness, we provide the DM expression for $\langle\hat{\mathcal{P}}_{2k} \rangle$
\begin{equation}\label{dmp2k}
\langle\hat{\mathcal{P}}_{2k} \rangle=(-1)^{k}\left(\frac{1}{k!}\right)^2\sum _{\substack{p_1,p_2,\ldots,p_k\\q_1,q_2,\ldots,q_k}}\sum _{\substack{r_1,r_2,\ldots,r_k\\s_1,s_2,\ldots,s_k}}\left[\prod_{i=1}^kS^{s_i}_{p_i}S^{q_i}_{r_i}\right]\left(\Gamma_{A}^{(k)}\right)^{p_1p_2\ldots p_k}_{q_1q_2\ldots q_k}\left(\Gamma_{B}^{(k)}\right)^{r_1r_2\ldots r_k}_{s_1s_2 \ldots s_k}.
\end{equation}
\twocolumngrid
Equations~\eqref{intermed}-\eqref{cinter} and \eqref{dmp2k} reveal the numerical complexity of the problem: calculation of the first-order exchange energy through the $2n$-order in the overlap expansion requires $k$-RDMs with $k$ up to $n+1$. In particular, the double-exchange approximation to the first-order exchange energy, defined previously as the sum of Eqs.~\eqref{Es2} and \eqref{Es4}, involves $k$-RDMs with $k=1,2,3$. Thus, the DM approach can be applied with any multireference method that provides these quantities. Storing and multiplying such objects quickly becomes computationally demanding. Therefore, it becomes crucial to take advantage of the factorization of the density matrices \cite{McWeeny:59}. For example, in SAPT based on Hartree-Fock wave functions, the first-order exchange energy can be calculated to all orders in the intermonomer overlap.~\cite{Jeziorski:76} This is possible because, for a single Slater determinant, $k$-RDMs, $k>1$, can be expressed as products of 1-RDMs. In multideterminant group-product wave functions, such as DMRG or CASSCF, electron correlation is present, so second-, third-, and higher-order cumulants of the density matrix generally do not vanish \cite{Kutzelnigg:97,Kutzelnigg:99}. In the following, we consider only the special case of strongly orthogonal geminal wave functions, where one can take advantage of a particularly simple structure of the density matrix cumulants.

In the APSG theory \cite{Kutzelnigg:64,Surjan:99} we assume that a $2N$-electron wave function is built from geminals
\begin{equation}
   \ket{\Phi}=2^{-N}\hat{\mathcal{A}}\left[\prod_{I=1}^N g^{(I)}(2I-1,2I)\right] ,
\end{equation}
which are strongly orthogonal, i.e., for $K \neq L$:
\begin{equation}
  \int \text{d}1 \, g^{(K)}(1,2)g^{(L)}(1,3) = 0,
\end{equation}
where $1\equiv(\mathbf{r},\sigma)$ denotes combined spatial and spin coordinates. Each geminal is expanded into a set of natural orbitals $\varphi_i^K(\mathbf{r})$:
\begin{equation}
g^{(K)}(1,2) = \sum_{i=1}^{N_K}c_{i}^K\varphi^{K}_i(\mathbf{r}_1)\varphi^K_i(\mathbf{r}_2) \chi(\sigma_1,\sigma_2),
\end{equation}
where $\chi$ is a two-electron singlet spin function. The geminal is normalized according to $\sum_{i=1}^{N_K}(c_i^K)^2=1$, where $N_K$ denotes the number of orbitals in geminal $K$. In the special case of a GVB-PP wave function $N_K=2$ for all $K$.
Geminal expansion coefficients are related to the occupation numbers $n^K$ by the relation $n^K_i=(c^K_i)^2$. In view of the Arai theorem \cite{Arai:60}, orbitals belonging to different geminals are orthogonal
\begin{equation}
    \int \text{d}\mathbf{r} \, \varphi^K_i(\mathbf{r}) \varphi^L_j(\mathbf{r}) = \delta_{KL} \delta_{ij}.
\end{equation}
In the second-quantized form, we can represent the $K$th geminal as
\begin{equation}
\begin{split}
 \hat{G}(K)\ket{0} &= \sum_{i=1}^{N_K} c^K_i x^\dagger_{i_K \alpha} x^\dagger_{i_K \beta} \ket{0} \\
  &= \frac{1}{2}\sum_{i=1}^{N_K} c^K_i (x^\dagger_{i_K \alpha} x^\dagger_{i_K \beta} 
 - x^\dagger_{i_K \beta}x^\dagger_{i_K \alpha}) \ket{0} ,
\end{split}
\end{equation}
where $\ket{0}$ is the physical vacuum state and $x^\dagger_{i_K \alpha}$ and $x^\dagger_{i_K \beta}$ denote creation operators associated with the $\varphi^{K}_{i}(\mathbf{r})\alpha(\sigma)$ and $\varphi^{K}_{i}(\mathbf{r})\beta(\sigma)$ spinorbitals, respectively.

In Appendix A, we provide general DM expressions for the expectation values $\braket{\hat{V}\hat{\mathcal{P}}_4}$ and $\braket{\hat{\mathcal{P}}_4}$ for closed-shell systems, which involve up to three-electron RDMs. For APSG/GVB-PP wave functions, these expressions can be evaluated using the spin-resolved 2- and 3-RDMs presented in Appendix B. Alternatively, for geminal wave functions, one can directly (i.e., without the use $k$-RDMs) compute expectation values using a diagrammatic technique originally developed by Paldus and co-workers~\cite{Paldus72I,Paldus72II} and adapted here to tensor products of monomer Fock spaces. In practice, we followed the diagrammatic route, as it proved more convenient both for derivation and for subsequent implementation: translating diagrams into algebraic code can be easily automated. Since the final expression involves more than 120 terms, we present it in the supplementary material along with additional details of the diagrammatic formulation. 

The direct computation of the first-order energy in the double-exchange approximation, possible for geminal wave functions, cannot be extended to arbitrary group-product functions \cite{McWeeny:59,McWeeny:60,Rosta:02}. Nevertheless, we can employ an approximate mapping between GVB-PP and CAS($n,n$)SCF wave functions, where CAS($m,n$) refers to the active space of $m$ electrons distributed on $n$ orbitals. This follows from the similarity between CAS($n,n$)SCF and GVB-PP two-electron RDMs, first observed by Baerends and co-workers \cite{vanMeer:18}. In practice, the mapping identifies strongly and weakly occupied CASSCF orbitals that effectively pair into geminals. The identification of proper pairings follows from comparing elements of the 2-RDM from a CAS($n$,$n$)SCF calculation with the 2-RDM constructed from CASSCF occupation numbers according to the GVB-PP formula. Once approximate CASSCF geminals are found, double-exchange ($\propto S^4$) terms are evaluated directly from geminal expansion coefficients and overlap integrals. In this way, one avoids the necessity of computing and storing monomer 3-RDMs from CASSCF calculations.

Evaluation of the general DM-based expression for $E^{(1)}_{\rm exch}(\propto S^4)$ involves terms the computational cost of which scales as 8th power of the number of active orbitals. For strongly-orthogonal geminal wave functions, factorization of the 3-RDMs reduces the scaling to 6th power in the number of active geminal orbitals, defined here as the strongly occupied orbitals and their weakly occupied partners forming individual geminals.

\section{Computational details \label{sec:cd}}

The DM formulation of the $E^{(1)}_{\rm exch}(\propto S^4)$ energy correction was implemented in the GammCor \cite{gammcor} program. Monomer 1- and 2-RDMs for CASSCF, APSG and GVB-PP wave functions were obtained using a developer branch of Dalton \cite{Dalton:14}. Initial geminals for APSG/GVB-PP calculations were generated according to the protocol described in Ref.~\cite{Hapka:22}.

The reference first-order exchange energies at the FCI level of theory for few-electron systems (\ce{H2}$\cdots$\ce{H2}, \ce{He}$\cdots$\ce{H2}, and Be$\cdots$Be) were obtained using an in-house SAPT code. For the Be$\cdots$Be interaction, the calculations were performed with a frozen-core approximation similar to the index-range restriction (IRR) approach described by Patkowski and Szalewicz \cite{Patkowski:07}. Our SAPT(FCI) implementation employs a spin-free approach and an exact projection onto the [$2^2$] irreducible representation of the symmetric $S_4$ group which corresponds to the physical ground state of the four-electron dimer \cite{Korona:97,Patkowski:01}. Further details can be found in the supplementary material.

The reference SAPT(DFT) and SAPT0 first-order exchange energies for the water dimer were obtained using Molpro \cite{Molpro:12}. In DFT calculations, we used the PBE0 functional \cite{Perdew:96,Adamo:99} with the gradient-regulated asymptotic correction scheme \cite{Gruning:01}. First-order energies in the $S^4$ approximation for Hartree-Fock wave functions were computed in GammCor and tested against an independent implementation in the Psi4 \cite{Psi4:20} package. For the benzene-methanethial (\ce{CH2S}, also known as thioformaldehyde) dimer we performed reference supermolecular coupled-cluster (CC) calculations to establish positions of the minima. For the ground state, we employed the LNO-CCSD(T) approach \cite{Nagy:18,Nagy:19}, while excited-state interaction energies were obtained by shifting the ground-state LNO-CCSD(T) interaction energy by 
$\omega_{\rm CC2} = \omega^{A^{*}B}_{\rm CC2} - \omega^{A^{*}}_{\rm CC2}$, where $\omega^{A^{*}B}_{\rm CC2}$ and $\omega^{A^{*}}_{\rm CC2}$ are dimer and monomer, respectively, excitation energies computed at the CC2\cite{Christiansen:95} level of theory in the dimer basis set. All CC calculations were performed in the MRCC program \cite{mrcc,Kallay:20,Mester:25}.

For three systems, we employed the approximate mapping from CAS($n,n$) occupation numbers to geminals in calculations of $E^{(1)}_{\rm exch}(\propto S^4)$ terms (see the previous section). First, this approach was used to access the ${}^1\Sigma_u^+$ state of the hydrogen molecule in the He$\cdots$\ce{H2} complex. The second case was the calculation of the two lowest-lying singlet states of the \ce{CH2S} molecule. Finally, the mapping was applied to approximate double-exchange energy terms for CAS(8,8) monomer wave functions in the water dimer.

All calculations employed augmented correlation-consistent orbital basis sets of a triple-zeta quality (aug-cc-pVTZ) \cite{Dunning:89,Kendall:92} with monomers described using the dimer-centered basis set. For the beryllium dimer, we present results obtained in the smaller aug-cc-pVDZ basis, where it is feasible to include all active orbitals in the CASSCF active space.

\section{Results and discussion}
\subsection{Model few-electron systems: \ce{H2}$\cdots$\ce{H2}, He$\cdots$\ce{H2}, and Be$\cdots$Be}

\begin{figure*}
\centering
\includegraphics[width=\textwidth]{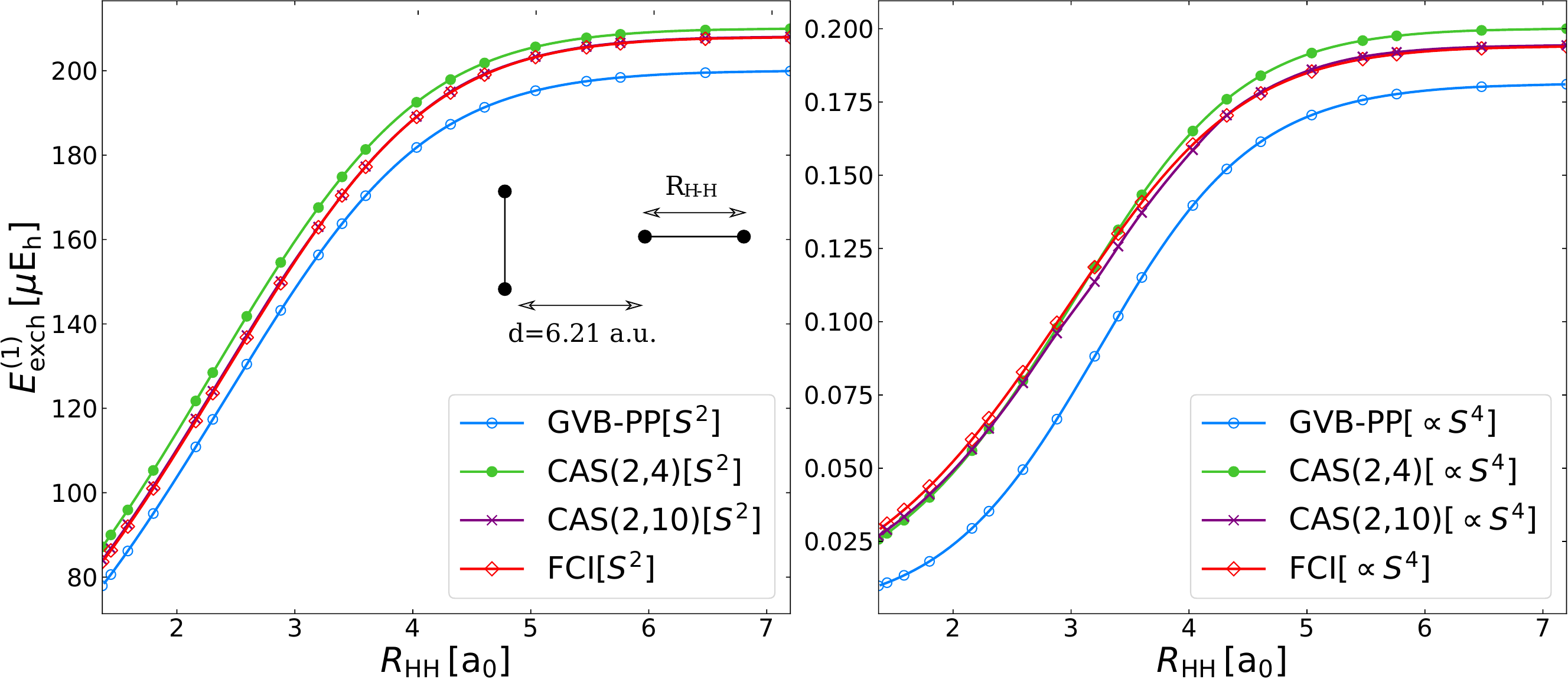}
\caption{First-order exchange energy terms proportional to the second ($S^2$) and the forth ($\propto S^4$) power of intermolecular overlap integrals computed at SAPT(FCI), SAPT(GVB-PP), and SAPT(CAS) levels of theory for the H$_2\cdots$H$_2$ dimer in a T-shaped geometry. The CAS($m,n$) notation refers to the active space on each monomer. The intermolecular distance is fixed at $6.21$~$a_0$. The $R_{\rm HH}$ label refers to the H-H bond length in one of the monomers which is varied from $1.37~a_0$ to $7.2~a_0$, while for the other monomer the H-H bond length is fixed at $1.44~a_0$.}
\label{fig:H2H2}
\end{figure*}

In this section, we discuss three model few-electron complexes where GVB-PP and APSG results can be directly compared with FCI reference values. As our first example, we consider the ground-state \ce{H2}$\cdots$\ce{H2} dimer in a T-shaped configuration, in which static correlation effects are gradually increased by stretching a single covalent bond in one of the monomers, while keeping the intermolecular distance fixed (see inset in Figure~\ref{fig:H2H2}).

Figure~\ref{fig:H2H2} compares first-order exchange energies obtained with GVB-PP, APSG, and FCI wave functions. Recall that for a two-electron system, APSG with $n$ weakly occupied orbitals is equivalent to a CAS(2,$n)$ function (the CAS notation is used in Figure~\ref{fig:H2H2}). We analyze exchange terms proportional to $S^2$ and $S^4$ separately as functions of the H-H bond length, $R_{\text{HH}}$. The GVB-PP-based description is already fairly accurate: the exchange curves remain nearly parallel to the FCI reference, with mean absolute errors of $-6.7$~$\mu$Ha and $-0.022$~$\mu$Ha in the $S^2$ and $\propto S^4$ components, respectively, corresponding to relative errors of $-4$\% and $-7$\% in the plateau ($R_{\rm HH}=5.8~a_0$) region. Increasing the number of orbitals in the geminal systematically improves the accuracy by capturing dynamic correlation effects.  In the stretched H-H bond regime, the errors in the $S^2$-approximated exchange for APSG with 4 and 10 correlating orbitals are $1$\% and 0.05\%, respectively. At the $S^4$ level, the errors drop from 3\% to 0.2\% when increasing the geminal space from 4 to 10 orbitals.

\begin{figure*}
\centering
\includegraphics[width=\textwidth]{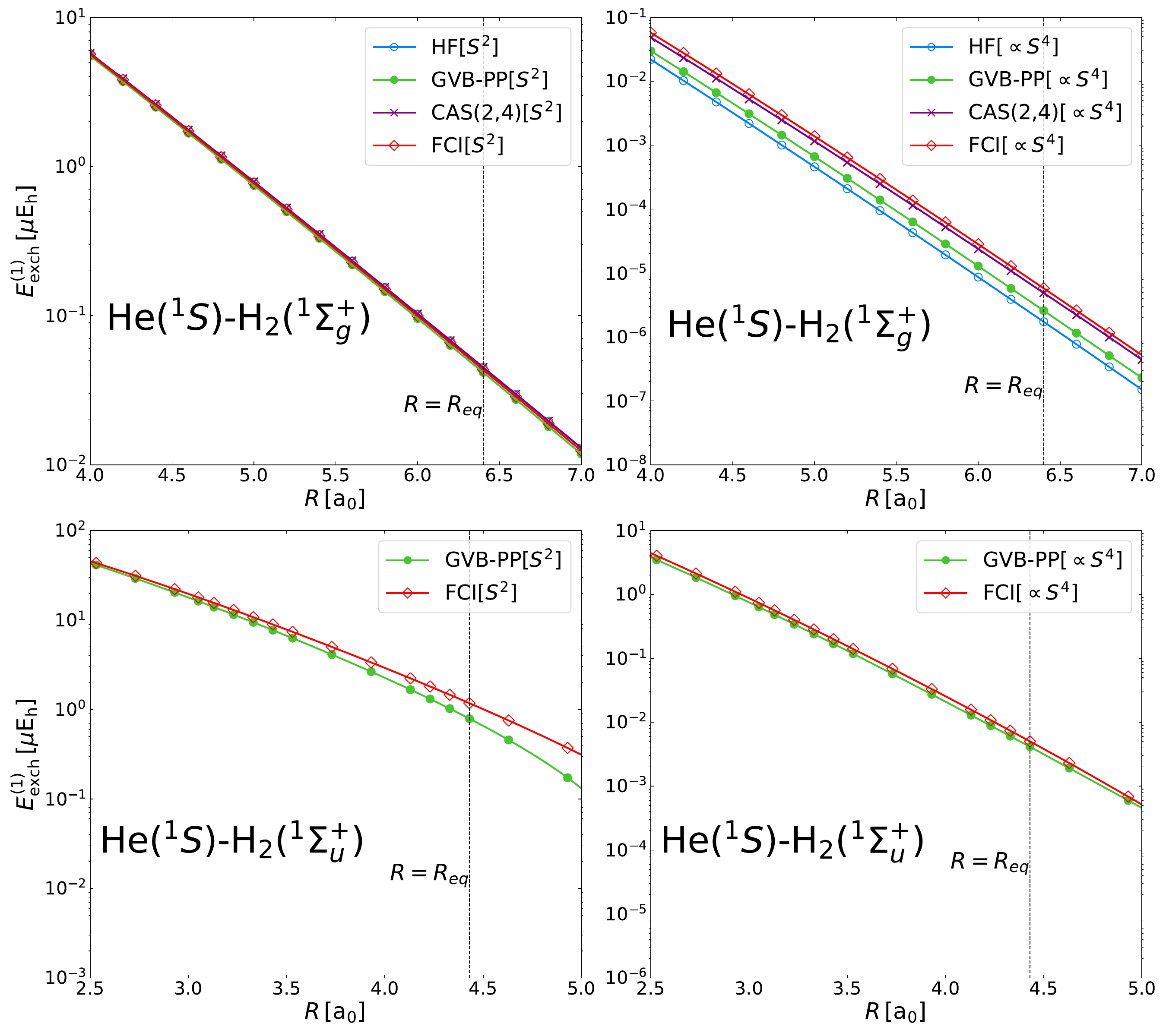}
\caption{First-order exchange energy components in the  $S^2$ approximation (left panels) and double-exchange contributions ($\propto S^4$, right panels) for the He$\cdots$\ce{H2} dimer in the T-shaped geometry. Top panels correspond to the ground He(${}^1$S)$\cdots$\ce{H2}(${}^1\Sigma_g^+$) state, bottom panels to the He(${}^1$S)$\cdots$\ce{H2}(${}^1\Sigma_u^+$) excited state. The HF, GVB-PP, CAS(2,4), and FCI labels denote SAPT(HF), SAPT(GVB-PP), SAPT(CAS), and SAPT(FCI) results. The $R=R_{\rm eq}$ vertical lines represent the equilibrium distance obtained from benchmark SAPT(FCI) calculations: $R_{\rm eq}=6.40~a_0$ for the ground state and $R_{\rm eq}=4.43~a_0$ for the excited state.}
\label{fig:HeH2}
\end{figure*}

As our second model system, we analyze the He$\cdots$H$_2$ dimer in a T-shaped arrangement. In this case, we examine two electronic states: the ground state, He(${}^1$S)$\cdots$\ce{H2}($^1\Sigma^+_g$), and the first singlet excited state, He(${}^1$S)$\cdots$\ce{H2}($^1\Sigma^+_u$). The equilibrium \ce{H2} bond length is 1.44~$a_0$ for the ground and 2.44~$a_0$ for the excited state. In addition to geminal-based and FCI results, we also report exchange curves for the ground state using the Hartree-Fock (HF) description of the monomers. Figure~\ref{fig:HeH2} presents the exchange energy in the $S^2$ approximation and the $\propto S^4$ contribution, plotted as functions of the distance between the He atom and the midpoint of the H-H bond.

For the He(${}^1$S)$\cdots$\ce{H2}(${}^1\Sigma_g^{+}$) ground state, all SAPT variants correctly recover single-exchange contribution, with relative errors not exceeding 5\% along the entire potential energy curve when compared to SAPT(FCI). The double-exchange term, however, is more sensitive to dynamic correlation effects. The HF-based SAPT recovers only between 30\% and 40\% of the $\propto S^4$ energy. The GVB-PP reference offers a slight improvement, reproducing 40-50\% of the FCI result. Extending the active space by two correlating orbitals per monomer is sufficient to reach 85\% accuracy, cf.\ the CAS(2,4) curve in Figure~\ref{fig:HeH2}.

The situation changes significantly for the He(${}^1$S)$\cdots$\ce{H2}(${}^1\Sigma_u^{+}$) excited state, where we compare GVB-PP and FCI-based results (see Figure~\ref{fig:HeH2}). Relative to the ground state, the van der Waals minimum shifts from 6.40~a$_0$ to 4.43~a$_0$, mainly due to the reduced exchange repulsion in the midbond plane of \ce{H2}(${}^1\Sigma_u^{+}$), which allows helium to approach closer. At short intermonomer separations, GVB-PP matches the FCI description of the single-exchange energy, but fails to capture the correct long-range decay. While the exact $S^2$ exchange energy becomes attractive for distances larger than 6.4~a$_0$, SAPT(GVB) predicts the sign change already at 5.4~a$_0$. The double exchange contribution is purely repulsive at all distances, and the corresponding GVB-PP and FCI curves remain nearly parallel and deviate by at most 18\%. 

\begin{figure*}
\centering
\includegraphics[width=0.98\textwidth]{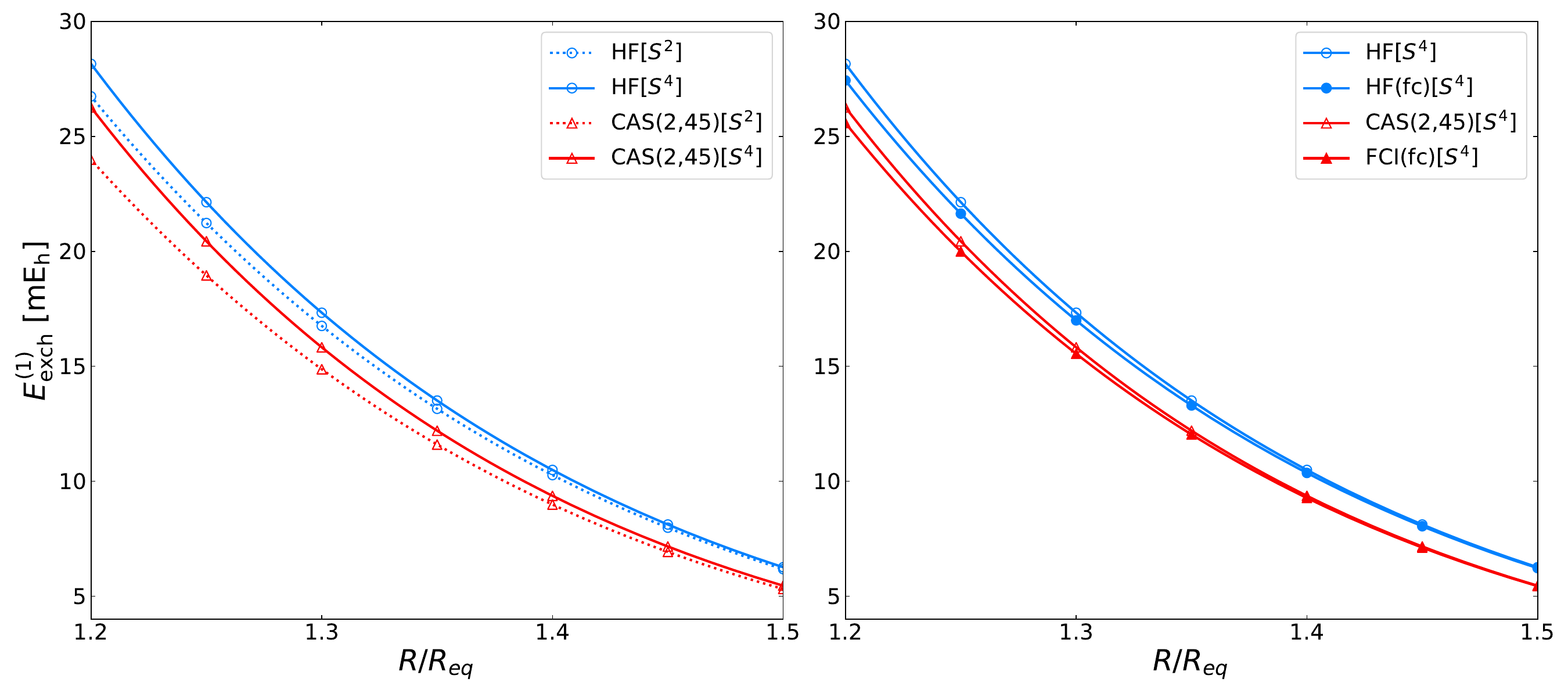}
\caption{First-order exchange energy in the Be$\cdots$Be dimer. Left panel: comparison of $S^2$ and $S^4$ approximations at SAPT(HF) and SAPT(CAS) levels of theory. Right panel: comparison of first-order exchange based on Hartree-Fock, CASSCF and FCI description of monomers. fc refers to the frozen-core approximation (see text for details). The equilibrium distance is 4.6~a$_0$. The basis set is aug-cc-pVDZ.}
\label{fig:BeBe}
\end{figure*}

Our final model system is the beryllium dimer. In their study of alkaline-earth-metal dimers, Patkowski et al.~\cite{Patkowski:07b} demonstrated that first-order exchange contributions beyond the $S^2$ approximation are particularly large for this complex. Near equilibrium, they amount to 9-11\% of the non-approximated result, as evaluated with single-reference SAPT. In the left panel of Figure~\ref{fig:BeBe}, we compare exchange energies computed with Hartree-Fock and CAS(2,45)SCF monomer wave functions, the latter including all orbitals in the active space except for the 1$s$ orbital which remains inactive. As can be seen, electron correlation reduces the overall repulsion. Errors in the $S^2$ and $\propto S^4$ terms partially cancel out: SAPT(HF) systematically overestimates the $S^2$ term, while underestimating the $\propto S^4$ contributions (at equilibrium, the respective errors are 13\% and $-25$\%). The majority of electron correlation can be recovered already with the minimal CAS(2,4) active space on Be: 98\% and 92\% of the reference $E^{(1)}_{\rm exch}(S^2)$ and $E^{(1)}_{\rm exch}(\propto S^4)$ values, respectively (see Table~S1 in the supplementary material).

Unlike exchange interactions between two-electron systems, the beryllium dimer involves nonzero 3-RDM contributions, allowing us to probe all terms in the double-exchange energy expression. This makes Be$\cdots$Be a more stringent test for our approach. The right panel in Figure~\ref{fig:BeBe} shows the exchange energy in the $S^4$ approximation (the sum of $S^2$ and $\propto S^4$ terms), computed using two methods: the formula developed in this work and SAPT(FCI) with the frozen-core (fc) approximation. At the HF level, the fc approach underestimates the all-electron results, with errors reaching 3\% at the $R$/$R_{\rm eq}$=1.2 monomer separation. The CAS(2,45) and FCI(fc) curves exhibit the same behavior, validating our implementation. At shorter distances, the fc approximation begins to break down. In fact, fc errors in the individual $S^2$ and $\propto S^4$ components systematically cancel---within the range plotted in Figure~\ref{fig:BeBe}, individual errors increase from 1\% to 12\% (see Table~S1 in the supplementary materials).

\subsection{Model many-electron systems: Be$\cdots$\ce{NH3}, \ce{C6H6}$\cdots$\ce{CH2S}, and \ce{H2O}$\cdots$\ce{H2O}}

The Be$\cdots$\ce{NH3} interaction is particularly challenging for approximate treatment of exchange effects. The global minimum occurs at a short distance of 3.6~a$_0$, with Be approaching the ammonia lone pair along the C$_3$ axis \cite{Zuchowski:08}. Chałasiński et al.~\cite{Chalbie:93} investigated this complex in the context of proton-donor properties of ammonia. Żuchowski and Hutson \cite{Zuchowski:08} later characterized its stationary points from the perspective of producing ultracold ammonia via sympathetic cooling. More recently, Tyrcha et al.~\cite{Tyrcha:24} demonstrated that effects beyond the $S^2$ approximation are particularly strong. Notably, the $\propto S^4$ contributions through second order in $\hat{V}$ are 12.5 times larger in magnitude than the total interaction energy.

\begin{figure*}
\centering
\includegraphics[width=0.9\textwidth]{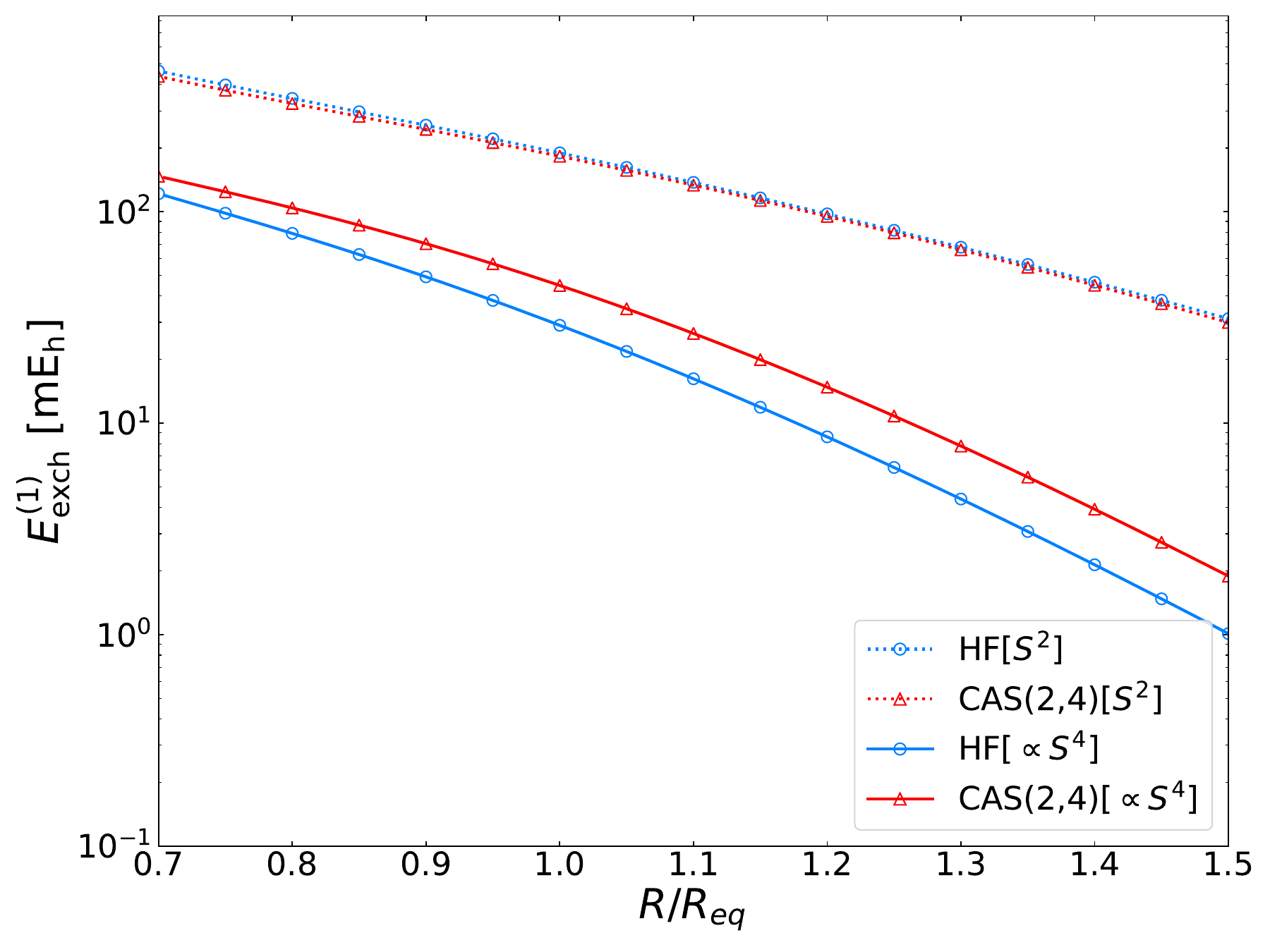}
\caption{First-order exchange energy in the Be$\cdots$\ce{NH3} dimer computed at SAPT(HF) and SAPT(CAS) levels of theory. The ``CAS(2,4)'' label denotes the active space on Be, while \ce{NH3} is described with a HF wave function. Geometry is taken from Ref.~\citenum{Tyrcha:24} and the basis set is aug-cc-pVTZ.}
\label{fig:BeNH3}
\end{figure*}

In Figure~\ref{fig:BeNH3}, we present the $E^{(1)}_{\rm exch}(S^2)$ and $E^{(1)}_{\rm exch}(\propto S^4)$ energy terms obtained from SAPT(HF) and SAPT(APSG) calculations as functions of the Be-N distance along the $C_3$ axis. To illustrate the impact of static correlation on the exchange interaction, we choose the minimal active space for SAPT(APSG), where beryllium is described with a single geminal constructed from the 2$s$ and 2$p$ orbitals (denoted as ``CAS(2,4)'' in Figure~\ref{fig:BeNH3}), while for ammonia we keep the HF reference.

In the $S^2$ approximation, the single- and multiconfigurational SAPT results differ by 3-6\%, with HF being more repulsive. In the case of double-exchange contributions, the difference is much more pronounced---SAPT(HF) underestimates APSG-based results by as much as 50\% in the long-range regime. As in the Be$\cdots$Be case, the overestimation of $S^2$ and underestimation of $\propto S^4$ terms leads to a partial cancellation of errors at the SAPT(HF) level. Consistent with Ref.\citenum{Tyrcha:24}, we find that the contribution of the $\propto S^4$ terms relative to $E^{(1)}_{\rm exch}(S^4)$ in the minimum amounts to 20\% and 13\% in SAPT(APSG) and SAPT(HF), respectively, which is extremely high for a neutral, closed-shell complex.

\begin{figure*}
\centering
\includegraphics[width=0.98\textwidth]{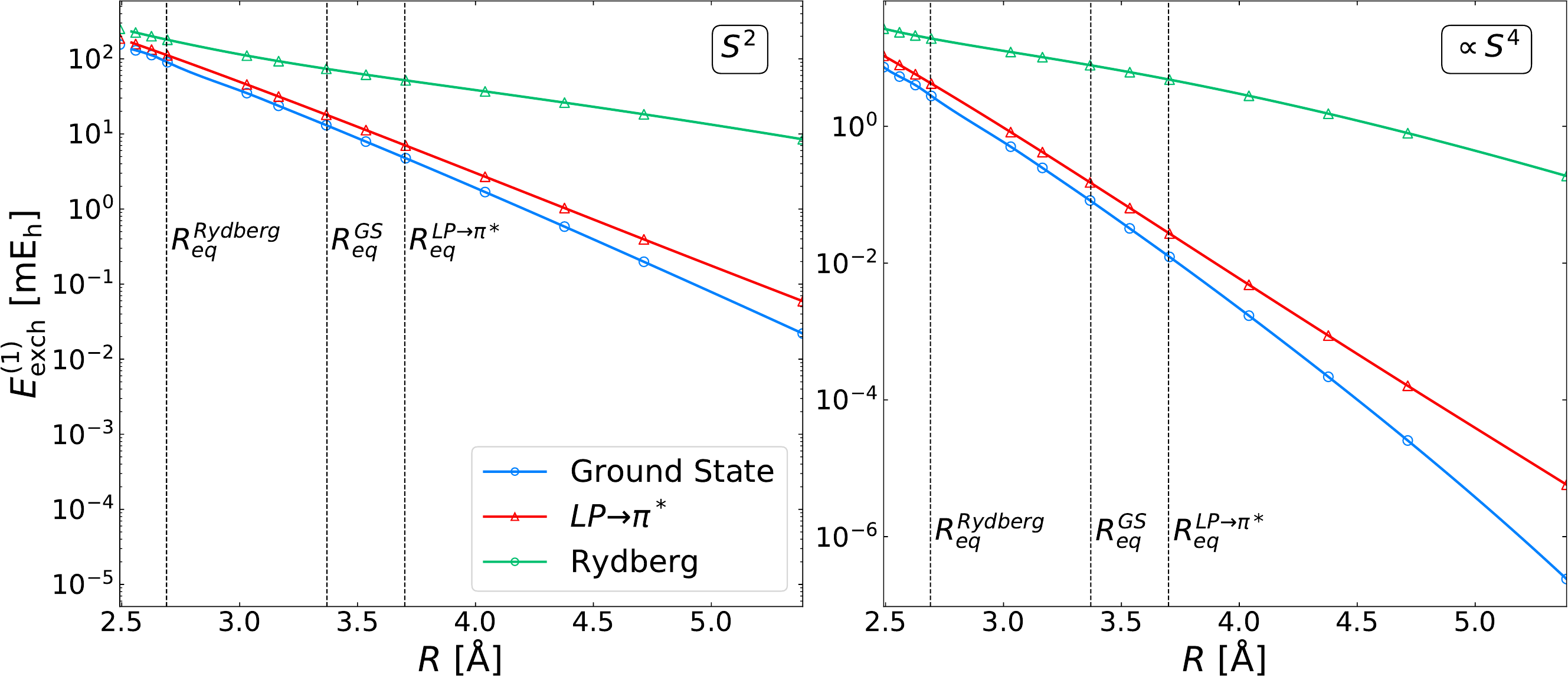}
\caption{First-order exchange energy in the \ce{C6H6}$\cdots$\ce{CH2S} dimer in stacked geometry computed at the SAPT(CAS) level of theory in the ground state and the two lowest vertically-excited singlet states: valence (LP$\to\pi^*$) and Rydberg. Benzene and methanethial are described using the Hartree-Fock and the CAS(2,3) wave functions, respectively. Vertical lines correspond to the positions of the minima from reference CC calculations (see text for details). The basis set is aug-cc-pVTZ.}
\label{fig:c6h6}
\end{figure*}

One of the main motivations for extending SAPT(MC) beyond the $S^2$ approximation are interactions involving excited-state molecules. Figure~\ref{fig:c6h6} shows first-order exchange energy contributions in the stacked benzene-methanethial (\ce{CH2S}) complex for the ground state and the two lowest singlet excited states, where the excitation is localized on \ce{CH2S}. This system exemplifies a chalcogen bond \cite{Mahmudov:17} which, in the stacked geometry, can be qualitatively interpreted as the interaction between the $\pi$-electron system on benzene and the positively charged region of \ce{CH2S}, localized on the C atom (the so-called $\pi$-hole) \cite{Wang:16,Sedlak:18}. The first electronic excited state of methanethial is dominated by a valence transition from the sulfur HOMO orbital of the lone pair (LP) character to the antibonding $\pi^*$ orbital on the C-S bond. The second excited state involves transition from the sulfur lone-pair to the LUMO+1 orbital (8~$a_1$ in an unperturbed molecule of the C$_{2v}$ symmetry) and has a Rydberg character.

The ground-state equilibrium geometry is taken from Ref.~\citenum{Sedlak:18}. The potential energy curves in Figure~\ref{fig:c6h6} are obtained by displacing the monomers along the intermolecular axis defined by their center-of-mass separation at equilibrium. All presented excited-state results correspond to vertical excitations.

Vertical excitations of methanethial in the stacked \ce{C6H6}$\cdots$\ce{CH2S} complex lead to enhanced exchange interactions. The effect is sizeable even in the valence LP$\to\pi^*$ state: the $S^2$ and $\propto S^4$ components increase by 40\% and 80\%, respectively, at the ground-state equilibrium geometry. The orbital-overlap interpretation is clear, as the sulfur lone-pair is oriented parallel to benzene, while the $\pi^*$ orbital is perpendicular to the benzene $\pi$ system. Although double-exchange contributions remain below 1\% of the total exchange energy, vertical excitation transfers the system from the ground state minimum to the repulsive wall of the LP$\to\pi^*$ potential, where higher-order contributions in the overlap expansion become relevant. For the Rydberg excitation, the exchange enhancement is drastic, with the $S^2$ and $\propto S^4$ terms increasing by one and two orders of magnitude, respectively. Double-exchange effects become significant, contributing about 10\% to the total first-order exchange energy at the ground-state equilibrium geometry, which maps to the attractive region of the Rydberg potential (cf.\ the position of the Rydberg minimum in Figure~\ref{fig:c6h6}).

\begin{figure*}
    \centering
    \includegraphics[width=1\linewidth]{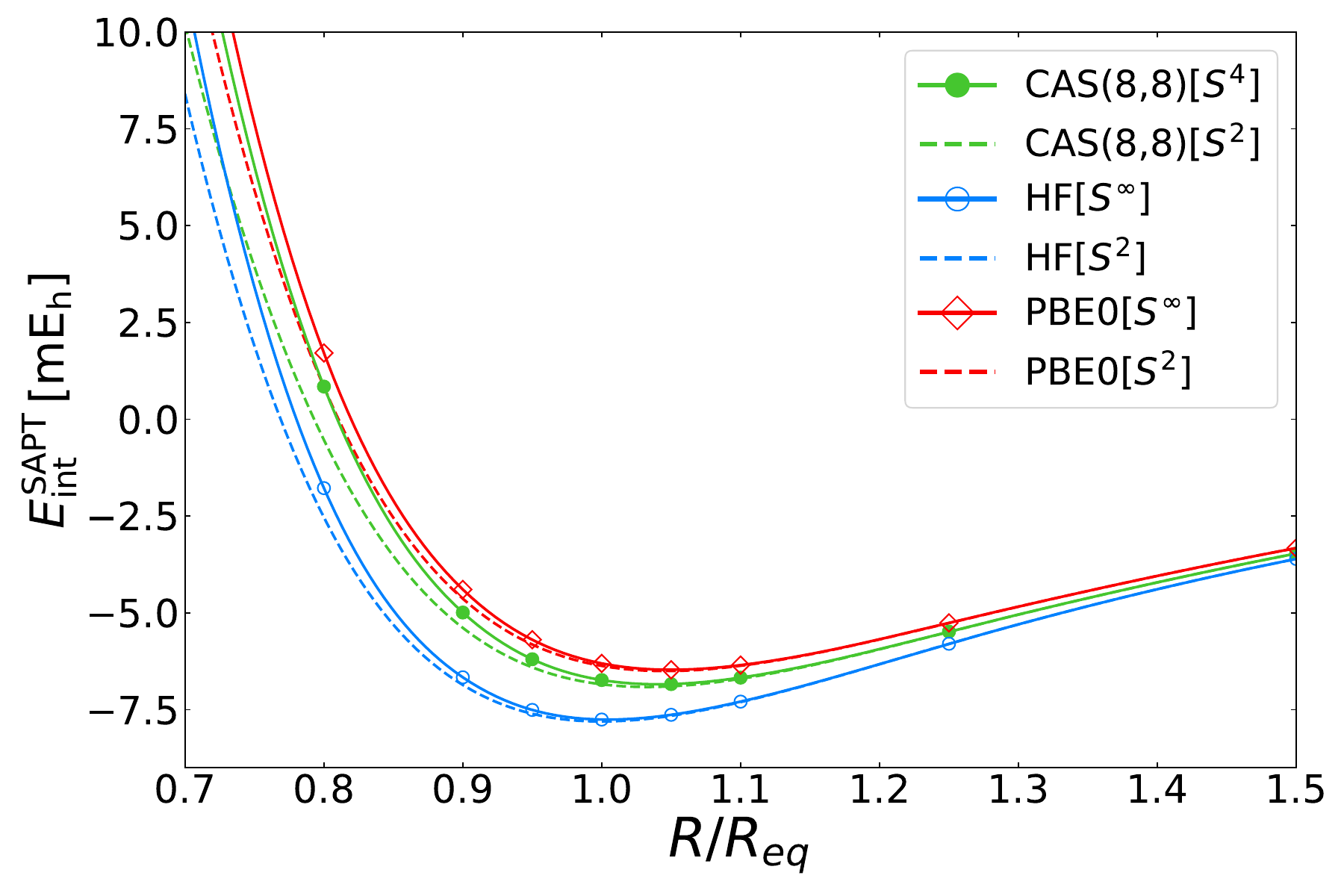}
    \caption{Interaction energy curves obtained with second-order SAPT approaches for the water dimer. Dashed lines represent results obtained with the $S^2$ approximation in the exchange components. Solid lines represent interaction energies obtained either with non-approximated first-order exchange [SAPT(HF) and SAPT(PBE0)], or in the double-exchange approximation [SAPT(CAS)]. In the second-order interaction energy terms, the $S^2$ approximation is used in all exchange components. Geometries were taken from the S66x10 dataset.~\cite{Smith:16} The basis set is aug-cc-pVTZ.}
\label{fig:h2o}
\end{figure*}

In our final example, we illustrate how the first-order double-exchange ($\propto S^4$) terms improve the short-range part of the SAPT(MC) potential. Figure~\ref{fig:h2o} presents interaction energy curves for the water dimer from SAPT(HF), SAPT(PBE0), and SAPT(CAS) calculations. For consistency, all SAPT variants include second-order exchange terms limited to the $S^2$ approximation. As expected, the single-exchange approximation in the first order remains valid from the long-range regime down to the potential minimum. While SAPT(HF) underestimates the $E^{(1)}_{\rm exch}(S^2)$ term by 7-21\% relative to SAPT(PBE0), SAPT(CAS) recovers 96-99\% of the DFT-based reference. In the repulsive region, accounting for first-order double-exchange contributions improves the agreement between SAPT(CAS) and reference SAPT(PBE0) potentials (denoted as $S^4$ and $S^\infty$, respectively, in Figure~\ref{fig:h2o}). A closer inspection reveals that the individual $\propto S^4$ terms from SAPT(CAS) are about 1.5-2.0 times larger than their single-reference counterparts, similar to the Be$\cdots$\ce{NH3} case.

\section{Conclusions}

We have presented an extension of the SAPT(MC) method \cite{Hapka:19b,Hapka:21} to account for double-exchange ($\propto S^4$) effects in the first-order interaction energy. This was accomplished by generalizing the DM approach of Moszynski et al.~\cite{Moszynski1994a} to arbitrary orders in the overlap expansion. Evaluating the first-order exchange energy in the $S^4$ approximation requires access to 1-, 2-, and 3-RDMs of the monomers.

The $E^{(1)}_{\rm exch}(S^4)$ model was implemented for strongly orthogonal wave functions of the APSG and GVB-PP types. The resulting geminal-based expressions can also approximate the $\propto S^4$ energy terms in SAPT(CAS) calculations by effectively pairing active orbitals from valence CAS($n,n$) wave functions into the GVB-PP-type geminals. The pairing scheme exploits the structural similarity between the 2-RDM of CAS($n,n$) and GVB-PP wave functions \cite{vanMeer:18,Hapka:20}.

We validated our SAPT(APSG) implementation against benchmark SAPT(FCI) results for model multireference systems. In small dimers composed of ``single-geminal'' monomers (\ce{H2}$\cdots$\ce{H2}, He$\cdots$\ce{H2}, Be$\cdots$Be), more than 85\% of intramonomer correlation in the $\propto S^4$ terms can be recovered by including only four orbitals in the active space. The Be$\cdots$Be and Be$\cdots$\ce{NH3} complexes exemplify a spectacular breakdown of the single-exchange approximation: already at equilibrium, double-exchange contributes 16-20\% of the total first-order exchange. In these systems, the individual $S^2$ and $\propto S^4$ components from SAPT(HF) and SAPT(APSG) calculations differ significantly, reflecting the multireference character of the Be atom. Nevertheless, SAPT(HF) errors largely cancel out. A more general observation is that the $E^{(1)}_{\rm exch}(\propto S^4)$ terms evaluated with multireference wave functions are larger in magnitude than their single-reference counterparts. This trend holds for both strongly and weakly correlated molecules and can be rationalized by noting that the $\hat{\mathcal{P}}_4$ operator couples different excited determinants from the bra and ket states. These ``off-diagonal'' contributions are absent in the single-reference case.

To verify the significance of exchange effects beyond the $S^2$ approximation in complexes involving electronically excited states, we examined two lowest excited states of the benzene-methanethial dimer. As seen for the first valence LP$\to\pi^*$ excitation, vertical transition can drive the system into repulsive region of the potential. In the second excited state of a Rydberg character, the $S^4$ exchange increases by one order of magnitude with respect to the ground state, so that double-exchange terms become pronounced even in attractive regions. This typically signals the breakdown of the perturbation expansion in powers of $\hat{V}$ and necessitates a hybrid approach incorporating supermolecular methods \cite{Patkowski:07,Hapka:12,Tyrcha:24}.

Since the 1990s, the combination of first-order exchange energy accurate in the overlap expansion with second-order terms in the $S^2$ approximation has been the standard in SAPT \cite{Patkowski:20,Korona:23}. To this day, it remains the default choice in the most widely used implementations of the method. The present work brings SAPT(MC) closer to this benchmark. The development of unapproximated second-order \cite{Schaffer:12,Schaffer:13} and third-order exchange-induction contributions \cite{Waldrop:21} has further improved the accuracy of single-reference SAPT. Extending these advances to the multiconfigurational formulation of the method is an ongoing effort in our group.

\begin{acknowledgements}
The National Science Center of Poland supported this work under grant no.\ 2021/43/D/ST4/02762. 
We gratefully acknowledge Polish high-performance computing infrastructure PLGrid (HPC Center: ACK Cyfronet AGH) for providing computer facilities and support within computational grant no.\ PLG/2024/016964. For the purpose of Open Access, the author has applied a CC-BY public copyright license to any Author Accepted Manuscript (AAM) version arising from this submission.
\end{acknowledgements}

\section*{Supporting Information}
Details on the SAPT(FCI) implemention for two-electron monomers. Outline of the diagrammatic SAPT(APSG) approach, including all $E^{(1)}_{\rm exch}(S^4)$ diagrams and the corresponding algebraic expressions.

\section*{Appendix A} Here, we present the working expressions for calculating $\langle \hat{V} \hat{\mathcal{P}}_4 \rangle$ and $\langle \hat{\mathcal{P}}_4 \rangle$ in the DM formulation for a closed-shell system.
We evaluate $\langle\hat{\mathcal{P}}_4\rangle$ as:
\onecolumngrid
\begin{align*}
    \langle \hat{\mathcal{P}}_{4} \rangle &=\sum^{} _{\substack{p_1,p_2\\q_1,q_2}}\sum^{} _{\substack{r_1,r_2\\s_1,s_2}}S^{s_1}_{p_1}S^{s_2}_{p_2}S^{q_1}_{r_1}S^{q_2}_{r_2}\bigg(\frac{1}{2}\left(\Gamma_{A}^{(2)}\right)^{p_1^\alpha p_2^\alpha }_{q_1 ^\alpha q_2^\alpha }\left(\Gamma_{B}^{(2)}\right)^{r_1^\alpha r_2^\alpha }_{s_1^\alpha s_2^\alpha } + \left(\Gamma_{A}^{(2)}\right)^{p_1^\alpha p_2^\beta }_{q_1^\alpha q_2^\beta }\left(\Gamma_{B}^{(2)}\right)^{r_1^\alpha r_2^\beta }_{s_1^\alpha s_2^\beta } \bigg).
\end{align*}
The expression for $\langle \hat{V}\hat{\mathcal{P}}_{4} \rangle$ follows directly from Eq.~\eqref{dmvp2k} and reads
\begin{align*}
 \langle \hat{V}\hat{\mathcal{P}}_{4} \rangle =\left(\sum_{\substack{pq\in A\\rs \in B}} (G_2)^{pr}_{qs}\tilde{v}^{qs}_{pr} + \sum_{\substack{p\in A \\ rss_1 \in B}}(F_2)^{pr}_{s_1s}\tilde{v}^{s_1s}_{pr}+ \sum_{\substack{pqq_1\in A \\ r \in B}}(D_2)^{pr}_{qq_1}\tilde{v}^{qq_1}_{pr} +\sum_{\substack{pq_1\in A \\ rs_1 \in B}}(C_2)^{pr}_{s_1q_1}\tilde{v}^{s_1q_1}_{pr} \right),
\end{align*}
where the intermediates $G_2$, $F_2$, $C_2$, and $D_2$ are given as:
\begin{equation*}\label{G2eq}
\begin{split}
(G_2)^{p r }_{q  s}&= \frac{1}{2}\sum _{\substack{p_1,p_2\\q_1,q_2}}^{}\sum_{\substack{r_1,r_2\\s_1,s_2}}^{}S^{s_1}_{p_1}S^{s_2}_{p_2}S^{q_1}_{r_1}S^{q_2}_{r_2} \bigg(\left(\Gamma_{A}^{(3)}\right)^{p^\alpha p_1^\alpha p_2^\alpha}_{q^\al q_1^\al q_2^\al }\left(\Gamma_{B}^{(3)}\right)^{r^\al r_1^\al r_2^\al }_{s^\al s_1^\al s_2^\al } + 4 \left(\Gamma_{A}^{(3)}\right)^{p^\alpha p_1^\alpha p_2^\be}_{q^\al q_1^\al q_2^\be}\left(\Gamma_{B}^{(3)}\right)^{r^\al r_1^\al r_2^\be }_{s^\al s_1^\al s_2^\be }  \\ &+ \left(\Gamma_{A}^{(3)}\right)^{p^\alpha p_1^\be p_2^\be}_{q^\al q_1^\be q_2^\be}\left(\Gamma_{B}^{(3)}\right)^{r^\al r_1^\be r_2 ^\be }_{s^\al s_1^\be s_2^\be } +  \left(\Gamma_{A}^{(3)}\right)^{p^\alpha p_1^\alpha p_2^\alpha}_{q^\al q_1^\al q_2^\al }\left(\Gamma_{B}^{(3)}\right)^{r^\al r_1^\be r_2^\be }_{s^\al s_1^\be s_2^\be } +  \left(\Gamma_{A}^{(3)}\right)^{p^\alpha p_1^\be p_2^\be}_{q^\al q_1^\be q_2^\be }\left(\Gamma_{B}^{(3)}\right)^{r^\al r_1^\al r_2^\al }_{s^\al s_1^\al s_2^\al } \\ &+ 4 \left(\Gamma_{A}^{(3)}\right)^{p^\alpha p_1^\be p_2^\al}_{q^\al q_1^\be q_2^\al}\left(\Gamma_{B}^{(3)}\right)^{r^\al r_1^\al r_2^\be }_{s^\al s_1^\al s_2^\be } \bigg) 
\end{split}
\end{equation*}
\begin{equation*}\label{F2eq}
\begin{split}
  (F_2)^{pr }_{s_1 s}&=\sum _{\substack{p_2\\q_1,q_2}}\sum _{\substack{r_1,r_2\\s_2}}S^{q_1}_{r_1}S^{s_2}_{p_2}S^{q_2}_{r_2} \bigg(\left(\Gamma_{A}^{(2)}\right)^{p^\al p_2^\al }_{q_1^\al q_2^\al }\left(\Gamma_{B}^{(3)}\right)^{r^\al r_1^\al r_2^\al }_{s^\al s_1^\al s_2^\al} + 2 \left(\Gamma_{A}^{(2)}\right)^{p^\al p_2^\be }_{q_1^\al q_2^\be }\left(\Gamma_{B}^{(3)}\right)^{r^\al r_1^\al r_2^\be }_{s^\al s_1^\al s_2^\be} \\ &+
\left(\Gamma_{A}^{(2)}\right)^{p^\al p_2^\al }_{q_1^\al q_2^\al }\left(\Gamma_{B}^{(3)}\right)^{r^\be r_1^\al r_2^\al }_{s^\be s_1^\al s_2^\al} +\left(\Gamma_{A}^{(2)}\right)^{p^\al p_2^\be }_{q_1^\al q_2^\be }\left(\Gamma_{B}^{(3)}\right)^{r^\al r_2^\al r_1^\be  }_{s^\al s_2^\al  s_1^\be } \bigg)
\end{split}
\end{equation*}
\begin{equation*}\label{D2eq}
\begin{split}
  (D_2)^{pr }_{q q_1 }&=\sum _{\substack{p_1,p_2\\q_2}}\sum _{\substack{r_2\\s_1,s_2}}S^{s_1}_{p_1}S^{s_2}_{p_2}S^{q_2}_{r_2} \bigg(\left(\Gamma_{A}^{(3)}\right)^{p^\al p_1^\al p_2^\al}_{ q^\al q_1^\al q_2^\al}\left(\Gamma_{B}^{(2)}\right)^{ r^\al r_2^\al }_{s_1^\al s_2^\al }+2\left(\Gamma_{A}^{(3)}\right)^{p^\al p_1^\al p_2^\be}_{ q^\al q_1^\al q_2^\be}\left(\Gamma_{B}^{(2)}\right)^{ r^\al r_2^\be }_{s_1^\al s_2^\be }
 \\ & + \left(\Gamma_{A}^{(3)}\right)^{p^\be p_1^\al p_2^\al}_{ q^\be q_1^\al q_2^\al}\left(\Gamma_{B}^{(2)}\right)^{ r^\al r_2^\al }_{s_1^\al s_2^\al }+ 2\left(\Gamma_{A}^{(3)}\right)^{p^\be p_1^\al p_2^\be}_{ q^\be q_1^\al q_2^\be}\left(\Gamma_{B}^{(2)}\right)^{ r^\al r_2^\be }_{s_1^\al s_2^\be }\bigg)
\end{split}
\end{equation*}
\begin{equation*}\label{C2eq}
\begin{split}
 (C_2)^{p   r }_{s_1 q_1 }&= 2\sum _{\substack{p_2\\q_2}}\sum _{\substack{r_2\\s_2}}S^{s_2}_{p_2}S^{q_2}_{r_2} \bigg(\left(\Gamma_{A}^{(2)}\right)^{p^\al  p_2^\al}_{q_1^\al q_2^\al }\left(\Gamma_{B}^{(2)}\right)^{r^\al r_2^\al}_{s_1^\al s_2^\al}+\left(\Gamma_{A}^{(2)}\right)^{p^\al  p_2^\be}_{q_1^\al q_2^\be }\left(\Gamma_{B}^{(2)}\right)^{r^\al r_2^\be}_{s_1^\al s_2^\be} +\left(\Gamma_{A}^{(2)}\right)^{p^\al  p_2^\be}_{q_1^\be q_2^\al  }\left(\Gamma_{B}^{(2)}\right)^{r^\al  r_2^\be }_{s_1^\be  s_2^\al } \bigg)
\end{split}
\end{equation*}
\section*{Appendix B}
We present the working expressions for computing 2- and 3-body RDMs corresponding to APSG/GVB-PP wave functions. Spin-separated components of the 2-RDM read 
\begin{equation*} 
\big(\Gamma^{(2)}_X\big)_{ q_1^\alpha q_2^\alpha}^{p_1^\alpha p_2^\alpha}=n^{I_{p_1}}_{p_1}n^{I_{p_2}}_{p_2}(\dk{p_1}{q_1}\dk{p_2}{q_2}-\dk{p_1}{q_2}\dk{q_1}{p_2})(1-\delta_{I_{p_1}I_{p_2}})  
\end{equation*}
and
\begin{equation*}
\big(\Gamma^{(2)}_X\big)_{ q_1^\alpha q_2^\beta}^{p_1^\alpha p_2 ^\beta} = \dk{p_1}{p_2}\dk{q_1}{q_2}\delta_{I_{p_1}I_{q_1}}c_p^{I_{p_1}}c_q^{I_{q_1}}
+ n^{I_{p_1}}_{p_1}n^{I_{p_2}}_{p_2}\dk {p_1}{q_1}\dk{p_2}{q_2}(1-\delta_{I_{p_1}I_{p_2}}),
\end{equation*}
where $I_p$ denotes the index of the geminal with contains the orbital $\varphi_p$. The spin-resolved 3-RDMs are obtained from the following expressions
\begin{align*}
\big(\Gamma^{(3)}_X\big)^{p_1^\al p_2^\al p_3^\al }_{q_1^\al q_2^\al q_3^\al} &= (1-\delta_{I_{p_1}I_{p_2}})(1-\delta_{I_{p_2}I_{p_3}})(1-\delta_{I_{p_3}I_{p_1}}) n_{p_1}n_{p_2}n_{p_3}
\bigg(\dk{p_1}{q_1}\dk{p_2}{q_2}\dk{p_3}{q_3}+\dk{p_1}{q_2}\dk{p_2}{q_3}\dk{p_3}{q_1} + \dk{p_1}{q_3}\dk{p_2}{q_1}\dk{p_3}{q_2}  \\ \nonumber
&- \dk{p_1}{q_2}\dk{p_2}{q_1}\dk{p_3}{q_3}-\dk{p_1}{q_3}\dk{p_2}{q_2}\dk{p_3}{q_1} - \dk{p_1}{q_1}\dk{p_2}{q_3}\dk{p_3}{q_2} \bigg)
\end{align*}
and
\begin{equation*}
\begin{split}
\big(\Gamma^{(3)}_X\big)^{p_1^\al p_2^\al p_3^\be }_{q_1^\al q_2^\al q_3^\be} &= (1-\delta_{I_{p_1}I_{p_2}}) n_{p_1}c_{p_2}c_{q_2} \delta_{I_{p_2}I_{q_2}}\dk{p_1}{q_1}\dk{p_2}{p_3}\dk{q_2}{q_3} 
+ (1-\delta_{I_{p_1}I_{p_2}}) n_{p_2}c_{p_1}c_{q_1}\delta_{I_{p_1}I_{q_1}}\dk{p_2}{q_2}\dk{p_1}{p_3}\dk{q_1}{q_3} \\
& - (1-\delta_{I_{p_1}I_{p_2}}) n_{p_1}c_{p_3}c_{q_3}\delta_{I_{p_3}I_{q_3}}\dk{p_1}{q_2}\dk{p_2}{p_3}\dk{q_1}{q_3} 
- (1-\delta_{I_{p_1}I_{p_2}}) n_{p_2}c_{p_3}c_{q_3}\delta_{I_{p_3}I_{q_3}}\dk{p_1}{p_3}\dk{p_2}{q_1}\dk{q_2}{q_3} \\
&+(1-\delta_{I_{p_1}I_{p_2}})(1-\delta_{I_{p_2}I_{p_3}})(1-\delta_{I_{p_3}I_{p_1}})n_{p_1}n_{p_2}n_{p_3}\dk{p_1}{q_1}\dk{p_2}{q_2}\dk{p_3}{q_3} \\
&-(1-\delta_{I_{p_1}I_{p_2}})(1-\delta_{I_{p_2}I_{p_3}})(1-\delta_{I_{p_3}I_{p_1}})n_{p_1}n_{p_2}n_{p_3}\dk{p_1}{q_2}\dk{p_2}{q_1}\dk{p_3}{q_3}
\end{split}
\end{equation*}
In Eqs.~\eqref{G2eq}--~\eqref{C2eq} it is convenient to use the 2-RDM symmetry
\begin{equation*}
\left(\Gamma_{X}^{(2)}\right)^{\kappa_1^\al  \kappa_2^\be}_{\lambda_1^\be \lambda_2^\al } = -  \left(\Gamma_{X}^{(2)}\right)^{\kappa_1^\al  \kappa_2^\be}_{\lambda_2^\al \lambda_1^\be}
\end{equation*}
and 3-RDM symmetry
\begin{equation*}
\left(\Gamma_{X}^{(3)}\right)^{\kappa_1^{\alpha}\kappa_2^{\alpha}\kappa_3^{\beta}}_{\lambda_1^{\alpha}\lambda_2^{\alpha}\lambda_3^{\beta}} = \left(\Gamma_{X}^{(3)}\right)^{\kappa_1^{\alpha}\kappa_3^{\beta}\kappa_2^{\alpha}}_{\lambda_1^{\alpha}\lambda_3^{\beta}\lambda_2^{\alpha}}= \left(\Gamma_{X}^{(3)}\right)^{\kappa_3^{\beta}\kappa_1^{\alpha}\kappa_2^{\alpha}}_{\lambda_3^{\beta}\lambda_1^{\alpha}\lambda_2^{\alpha}} =\left(\Gamma_{X}^{(3)}\right)^{\kappa_1^{\beta}\kappa_2^{\beta}\kappa_3^{\alpha}}_{\lambda_1^{\beta}\lambda_2^{\beta}\lambda_3^{\alpha}} 
= \left(\Gamma_{X}^{(3)}\right)^{\kappa_1^{\beta}\kappa_3^{\alpha}\kappa_2^{\beta}}_{\lambda_1^{\beta}\lambda_3^{\alpha}\lambda_2^{\beta}} 
= \left(\Gamma_{X}^{(3)}\right)^{\kappa_3^{\alpha}\kappa_1^{\beta}\kappa_2^{\beta}}_{\lambda_3^{\alpha}\lambda_1^{\beta}\lambda_2^{\beta}}.  
\end{equation*}
\twocolumngrid

%

\end{document}